\documentclass[superscriptaddress,twocolumn,aps,pre]{revtex4}
\usepackage{graphicx}
\begin{document}
\title{{\bf Evolutionary Dynamics and the Phase Structure of the Minority Game}}
\author{Baosheng Yuan}
\affiliation{Department of Computational Science, Faculty of Science, National University of Singapore, Singapore 117543}
\author{Kan Chen}
\affiliation{Department of Computational Science, Faculty of Science, National University of Singapore, Singapore 117543}

\date{\today }

\begin{abstract}

We show that a simple evolutionary scheme, when applied to the minority game (MG), changes the phase structure of the game. In this scheme each agent evolves individually whenever his wealth reaches the specified bankruptcy level, in contrast to the evolutionary schemes used in the previous works. We show that evolution greatly suppresses herding behavior, and it leads to better overall performance of the agents. Similar to the standard non-evolutionary MG, the dependence of the standard deviation $\sigma$  on the number of agents $N$ and the memory length $m$ can be characterized by a universal curve. We suggest a Crowd-Anticrowd theory for understanding the effect of evolution in the MG. 
\end{abstract}

\maketitle

{PACS numbers: 89.65.Gh, 87.23.Ge, 02.50.Le}

\section{Introduction}

Complex adaptive systems consist of agents using adaptive strategies to compete for limited resources. As changes in the global environment are induced by the agents themselves, it is important to study dynamics of such systems. The minority game (MG), proposed by Challet and Zhang\cite{challet1}, is a prototypical agent-based model that can be analyzed using the tools of statistical mechanics. The game captures some essential features of complex adaptive systems in which agents with limited information and rationality compete for limited resources. A key question in the study of agent-based models is, how evolution changes the behaviors of the agents.

There have been a few studies on the effect of evolution in the minority game. In the context of a simple evolutionary minority game, Johnson et al. found that the agents universally self-segregate into two opposing extreme groups \cite{johnson1}. Hod and Nakar, on the other hand, claimed that clustering of cautious agents emerges in a ``tough environment'' where the penalty for losing is greater than the reward for winning \cite{hod1}. Chen et al.~\cite{chen1,chen2} derived a general formalism to understand the dynamical mechanism for the transition from segregation to clustering. They found that the effective rate of evolution plays an important role in determining the resulting steady-state population distribution. These studies have focused mainly on population distribution.  Li et al.~\cite{li}, on the other hand, studied how evolution can help to improve the overall performance of the agents in the original MG. Starting from the adaptive MG proposed by Challet and Zhang \cite{challet1}, Li et al. introduced an evolutionary scheme in which all poorly performing agents evolve synchronously at every $\tau=10,000$ steps. Agents are ranked by their gains, and those ranked at the bottom \emph{p} percent (\emph{p}=10\%, 20\%, 30\%, 40\%, etc) are forced to change their strategies at these pre-specified steps. In order to make the evolution process smooth, not all the agents ranked at the bottom will change their strategies, but only 50\% of those (chosen randomly) have to do so. Those who are chosen to evolve replace the current strategies with new randomly picked ones. They reported that with evolution the  performance is significantly better; but the phase structure, characterized by the so called Savit curve \cite{savit}, remains similar to that of the original non-evolutionary MG. A later study \cite{yang} based on a variant of the evolutionary scheme used in Ref.~\cite{li} led to a similar conclusion, but with better overall performance of the agents.

When dealing with models of heterogeneous agent population, it makes sense to use an evolutionary scheme in which agents evolve individually, instead of synchronously at specified times. In this paper we adopt the simple scheme used in the EMG \cite{johnson1}, in which an agent becomes bankrupted and is replaced whenever its accumulated wealth is below a given threshold. With this simple scheme we found that herding behavior has disappeared when the memory length ($m$) is small, and the Savit curve obtained is significantly different from that of the original MG.

\section{Numerical Results}

Let us first briefly describe the minority game model. The game concerns a population of $N$ (odd number) heterogeneous agents with limited capabilities, who repeatedly compete to be in the minority group. After each round the winners gain a point and the losers lose a point. Each agent holds $S$ strategies. Each strategy is a look-up table listing the strategy's prediction of the minority group given the record of the most recent $m$ minority groups. There are total $2^{2^m}$ number of possible strategies, so the larger the value of $m$, the greater the processing power of the agents. Virtual points are accumulated for each of the strategies the agent has, and he uses the most successful strategy available to him. To include the effect of evolution we assign wealth $w$ to each agent; $w$ will increase/decrease by one when the agent wins/loses. The agent will be replaced if his wealth is below a threshold $-d$ ($d>0$); the new agent chooses his $S$ strategies randomly and his wealth is initialized to zero. The distribution of strategies gradually evolves as the game goes on.  

We have done extensive simulations with $N=51,101,201, 401$; $m=1,2,\cdots,10$; $S=1,2$; and $d=4,16,64,256,1024,4096$. The number of time steps used is set to be $T_0 = 800000 \times 2^{m/2} + 10000 d$; we have checked that this choice of the time step is sufficient for obtaining steady-state properties of the model. In our simulations we monitor $\sigma$, which is the standard deviation of the number of agents belonging to one of the groups. The smaller the value of $\sigma$, the larger a typical minority group and the better the overall performance of the agents will be. The smallest value of $\sigma$ is $\sigma= 0.5$, which means that the difference between the numbers in the minority and majority groups is one. The overall performance of the agents is at the optimal when $\sigma = 0.5$. At the other extreme, when the agents make their choices randomly as in the random choice game (RCG), we have $\sigma^2/N = 0.25$. In almost all our simulations, particularly for small $m$, evolution reduces $\sigma$ significantly. This is illustrated in Fig.~1, which shows $\sigma^2/N$ vs $2^m/N$ for the MG with and without evolution. The results are based on averages over eight independent runs, which are enough for obtaining accurate averages for the evolutionary MG, as the differences among different runs are quite small. For small $m$ evolution leads to a dramatic reduction in $\sigma$ as herding behavior of the agents is greatly suppressed. The results for different $N$ fall to a universal curve. For large $m$ the game still approaches the limit corresponding to the random choice game and  the effect of evolution is small.

\begin{figure}
\includegraphics*[width=8.5cm]{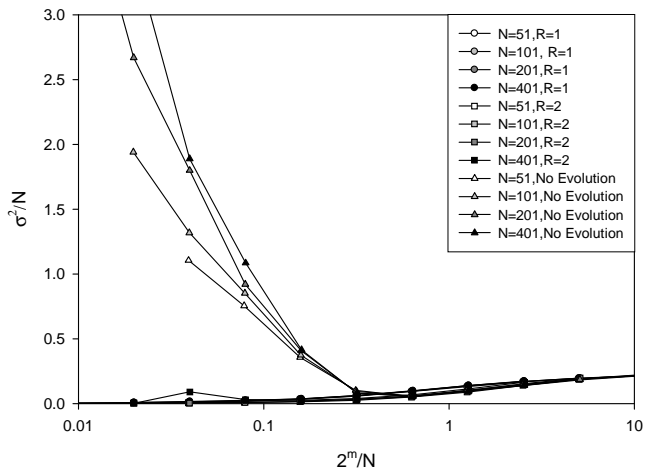}
\caption{ $\sigma^2/N$ vs $2^m/N$ for the MG with and without evolution. $d=256$ is used for the evolutionary MG. The results are obtained by averaging over eight independent runs}
\end{figure}

We have also studied the model with the award-to-fine ratio (as defined in Ref.~\cite{hod1}) $R \neq 1$. In this case the winners get $R$ points while the losers lose a point. We found that, for each $N$, there is an optimal value of $R = R_c(N) > 1$ that gives rise to the smallest value of $\sigma$. For $R>R_c(N)$ the average wealth is ever increasing and there is no steady state. Fig.~2 shows $\sigma^2/N$ vs $2^m/N$ for $R>1$ as compared to the case $R=1$. It is clear from the figure that $\sigma$ can be further reduced when $R>1$. We also found that, for $m=1$ or $2$, the optimal value of $\sigma=0.5$ can in fact be achieved.

\begin{figure}
\includegraphics*[width=8.5cm]{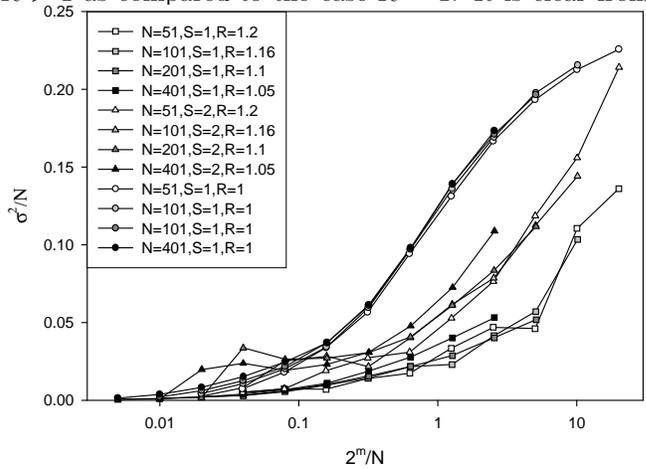}
\caption{ $\sigma^2/N$ vs $2^m/N$ for the evolutionary MG. $d=64$ is used. The results are obtained by averaging over eight independent runs}
\end{figure}

\section{A Crowd-Anticrowd theory for the evolutionary MG}

We now consider a Crowd-Anticrowd theory \cite{johnson2} to understand the effect of evolution in the minority game. Our discussion below follows Ref.~\cite{johnson2}. For simplicity we only consider $S=1$ and use a reduced strategy space (RSS). Numerically the differences between  the cases with $S=1$ and $S=2$ are small. RSS is a subset of strategies, which span the full strategy space (FSS). Consider, for example, an RSS for $m=2$, consisting of the following eight strategies:
\begin{widetext}
\begin{eqnarray}
U&\equiv&{\{-1 -1 -1 -1\}, \{+1 +1 -1 -1\}, \{+1 -1 +1 -1\}, \{-1 +1 +1 -1\}} \\
\bar{U} &\equiv & {\{+1 +1 +1 +1\}, \{-1 -1 +1 +1\}, \{-1 +1 -1 +1\}, \{+1 -1 -1 +1\}}
\end{eqnarray}
\end{widetext}
Here $\pm 1$ indicate the prediction of a strategy given one of the four possible histories. Any two strategies in \{$U$, $\bar{U}$\} are either uncorrelated (with the Hamming distance $2^m/2$) or anti-correlated (with the largest Hamming distance $2^m$). For a given $m$ there are total $2P$ ($P=2^m$) strategies with $P$ pairs of anti-correlated strategies (other pairs are uncorrelated) in RSS. For each strategy $G$ in RSS there is a corresponding anti-correlated strategy $\bar{G}$. It is believed that the essential features of the game are kept when RSS is used instead of the FSS \cite{challet2}.

Let us evaluate $\sigma^2=<(n_{+}(t) - N/2)^2>_t=<(n_{+}(t)-n_{-}(t))^2>/4$, where $n_+$ and $n_-$ are the numbers of agents making the choices $+1$ and $-1$ respectively. The average is over time step $t$. In RSS
$$n_+(t)-n_-(t) = \sum^{2P}_{G=1} a_G^{\mu(t)}n_G,$$
where $\mu(t)$ denotes the current history, $a_G^{\mu(t)} = \pm 1$ is the response of strategy $G$ to the history bit-string $\mu(t)$, and $n_G$ is the number of agents using strategy $G$ at time $t$. We have
\begin{equation}
\sigma^2 = \frac{1}{4}\sum^{2P}_{G=1,G'=1}<a_G^{\mu(t)}n_Ga_{G'}^{\mu(t)}n_{G'}>_t.
\end{equation}
In our simulation we found that the system visits all possible histories equally. This is illustrated in Fig.~3, which shows a histogram of the number of visits for each m-bit history. In order to compare with the non-evolutionary MG, we use $S=2$ in the figure; the result for $S=1$ is essentially the same. As we can see from the figure, the system does not visit all possible histories equally in the non-evolutionary MG.

\begin{figure}
\includegraphics*[width=8.5cm]{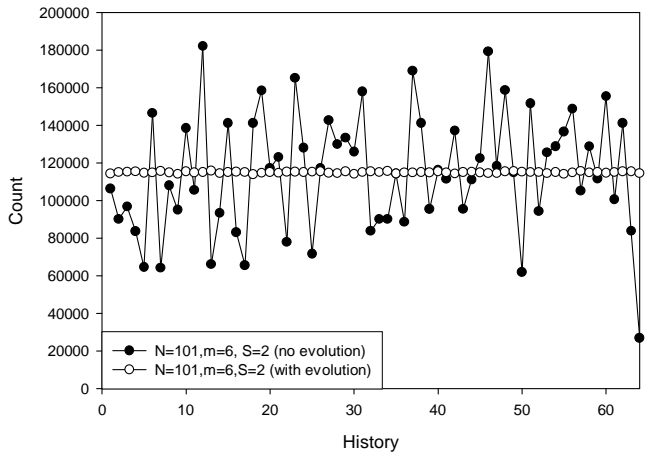}
\caption{Histogram for the number of appearances of all possible histories. $N=101$, $S=2$, $m=6$, 
and $d=256$. For comparison the corresponding histogram for the non-evolutionary MG is also plotted}
\end{figure}

For the evolutionary MG we can thus replace an average over time by an average over all possible histories. Now the double sum can be broken down into three parts, based on the correlation between the strategies. Given 
$<a_G a_G>_t = 1$, $<a_G a_{\bar{G}}>_t= -1$, and $<a_G a_{G'}>_t\approx <a_G a_{G'}>_h=0$ (for $G' \neq G$ and $G' \neq {\bar G}$), where $<\cdots>_h$ indicates average over all possible histories, we can write the double sum as
\begin{eqnarray}
\sigma^2 &=& \frac{1}{4}\left[\sum^{2P}_{G=1}<n_G^2>_t - \sum^{2P}_{G=1}<n_G n_{\bar{G}}>_t\right]\\
&=& \frac{1}{4}\sum^{P}_{G=1}\left[<n_G^2>_t - 2<n_G n_{\bar{G}}>_t +<n_{\bar{G}}^2>_t\right]\\
&=&\frac{1}{4}\sum^{P}_{G=1}<(n_G-n_{\bar{G}})^2>_t
\end{eqnarray}
In the above derivation we have tacitly assumed that $\{n_G\}$ change very slowly so the averages over $\{a_G\}$ can be done while holding $\{n_G\}$ constant. This is in the spirit of adiabatic approximation; it is valid because the evolution rate is low (as long as $d$ is not too small). 

Let $n_P$ be the number of pairs of agents holding anti-correlated strategies and $n_S$ be the number of agents holding unpaired strategies. Only the agents holding unpaired strategies contribute to $\sigma$; the game behaves essentially as an RCG with $n_S$ number of agents. Thus we have $\sigma^2 = 0.25 n_S$, or
\begin{equation}
\sigma^2/N = 0.25 s,
\end{equation}
where $s = n_S/ N$. Note that $2n_P + n_S=N$, $n_P$ can be written as $n_P = N(1-s)/2$. To determine $s$ we need to consider the evolutionary dynamics of the game. At the steady state we have the following balance equation
\begin{equation}
2r_P n_P(1-p(m,N)) = r_S n_S p(m, N),
\end{equation}
where $r_P$ and $r_S$ are the bankruptcy rate for the paired agents (pair breaking rate) and the bankruptcy rate for the unpaired agents respectively; $p(m,N)$ is the probability that a new pair is formed when a bankrupted agent is replaced. One can estimate $p(m,N)$ as
\begin{equation}
p(m, N)= 1- (1-1/2P)^{n_S} \approx 1 - \exp(-0.5Ns/2^m)
\end{equation}
It is somewhat difficult to estimate $r_P$ and $r_S$. It can be argued that $r_P/r_S$ does not  sensitively depend on $m$ and $N$. The equation for $s$ is then
\begin{equation}
\frac{r_P}{r_S}(1-s) = s \frac{p(s/z)}{1-p(s/z)},
\end{equation}
where $z=2^m/N$. The solution $s$ is a function of $z$. Thus, when we plot $\sigma^2/N$ vs $z=2^m/N$, the curves fall to a universal curve. Since $p(x)$ is a monotonically decreasing function of x, the universal curve $s(z)$ obtained from the above equation will be a monotonically increasing function of z. In the limit $z\rightarrow \infty$ we have $p(s/z) \rightarrow 0$; this leads to $s \rightarrow 1$ or $\sigma^2/N =0.25$. Thus $z\rightarrow \infty$ is the RCG limit. For $R>1$, $r_P/r_S$ decreases; this leads to smaller values for $s$ and $\sigma$. All these are in agreement with the simulation results. Thus the Crowd-Anticrowd picture provides a qualitative understanding of the evolutionary MG.

In conclusion, we have shown that, when a simple evolutionary scheme is applied to a heterogeneous population of agents, herding behavior in the MG is greatly suppressed.  The dependence of the standard deviation $\sigma$  on the number of agents $N$ and the memory length $m$ can be characterized by a universal curve. In addition, we demonstrated that a Crowd-Anticrowd theory can be used to understand qualitatively the effect of evolution in the MG.

This work was supported by the National University of Singapore research grant R-151-000-028-112.

\end{document}